\documentclass[%
 reprint,
nofootinbib,
 amsmath,amssymb,
 longbibliography,
]{revtex4-2}

\usepackage{graphicx}
\usepackage{dcolumn}
\usepackage{bm}

\usepackage{color}

\setcitestyle{numbers,square}

\begin{document}

\title{Spin interaction of non-relativistic neutrons with an ultrashort laser pulse}

\author{Peter Christian Aichelburg}
\affiliation{
 Fakult\"at f\"ur Physik, 
Universit\"at Wien}
\email{peter.christian.aichelburg@univie.ac.at}

\author{Christian Spreitzer}%
 \affiliation{Fakult\"at f\"ur Mathematik, 
Universit\"at Wien \& P\"adagogische Hochschule Nieder\"osterreich.}
\email{christian.spreitzer@univie.ac.at}

\begin{abstract}
The non-relativistic Pauli equation is used to study the interaction of slow neutrons with a short magnetic pulse. In the extreme limit, the pulse is acting on the magnetic moment of the neutron only at one instant of time.  We obtain the scattering amplitude by deriving the junction conditions for the Pauli wave function across the pulse. Explicit expressions  are given for a beam of polarized plane wave neutrons subjected to a pulse of spatially constant magnetic field strength. Assuming that the magnetic field is generated by an ultrashort laser pulse, we provide crude numerical estimates.
\end{abstract}

\maketitle





\section{Introduction}

The technique of generating ultrashort laser pulses offers many  new applications in  science and technology.  Lasers with femto- or even attosecond  pulses are able to produce  considerable  intensities reaching up to $10^{22}\,\mathrm{W/cm^2}$ \citep{Gaumnitz:17,Yo:19}. Such pulses allow to boost charged elementary particle to high energies, see e.g.\ \citet{CJ:22}, and may also be used to generate intense $\gamma$–rays and neutron beams via matter interaction  \citep{GU:22}. In this note we focus on the direct interaction of neutrons with an ultrashort magnetic pulse.  
  Assuming that the neutrons are non-relativistic,  the interaction is described by the Pauli equation.  The equation was first put forward  by W. Pauli  \citep{Pauli:27} to generalize the Schrödinger equation in order to study the interaction of charged spin-$\frac{1}{2}$ particles with external electromagnetic fields, but may equally describe the dynamics of neutral particles carrying a magnetic moment.\footnote{The Pauli equation can be derived as the non-relativistic limit of the relativistic Dirac equation which, in contrast to the Pauli equation, also models particles with high kinetic energy.}
Neutrons are uncharged and thus experience no Lorentz force,  hence the magnetic pulse 
  only acts on   their magnetic moment.   We model the physical pulse as the mathematical limit to act only at a single instant of time, i.e.\  as being proportional to $\delta (t)$.  This implies that before and after the interaction with the pulse, the neutrons satisfy the free Pauli equation and therefore the problem is to find solutions across the pulse.  The  $\delta$-pulse leads to mathematically ill-defined expressions, but these obstacles can be overcome by making use of the so called  Colombeau theory of generalized functions \citep{Colombeau:85,Colombeau:92}. 
 We will not expand on this here but merely  employ the result for our special case under consideration.  With this we are able to derive the matching conditions between the solution before and after the pulse.  For simplicity we specialize to plane wave neutrons. If the magnetic pulse is homogeneous, then the momentum of the wave remains unchanged, as expected, and only the spin orientation is affected by the interaction.  We explicitly derive the spin-flip probability as well as the spin expectation values for a beam of polarized neutrons. Finally we estimate the numerical outcome by making use of  technical data for high intensity ultrashort pulse lasers.

\section{The Pauli Equation for a $\delta$-like coupling term}\label{sec:paulidelta}
The Pauli equation for neutrons   is given by
\begin{equation}\label{pauliequ}
i \hbar \frac{\partial}{\partial t} \psi(t,\vec{x})= \left(
-\frac{\hbar^2}{2m} \, \triangle +\mu \,\, \vec{\sigma} \cdot
 \vec{\mathcal B}(t,\vec{x}) \right) \psi(t,\vec{x})
\end{equation}
where $\psi$ is a two-component wave function accounting for the particle's spin, $\vec{\mathcal B}(t,\vec{x})$ is the magnetic field strength,  
$\mu$ represents the absolute value of the magnetic moment of the neutron ($\mu= 0.966\,236\,45(24)\cdot 10^{-26}\, \mathrm{J\,T} ^{-1}$) and the sign takes care that the orientation is opposite to the spin. 
We model the magnetic pulse as a singular event in time 
 and set 
 \begin{equation}\label{deltapulse}
 \vec{\mathcal
B}(t,\vec{x})=\tau\,
\delta(t)\vec{B}(\vec{x}).
\end{equation}
At this stage, $\tau>0$ is  a not-yet specified constant introduced for dimensional reasons. It has the dimension of time, since $\delta(t)$ has the dimension of an inverse time. Later, when it comes to obtaining numerical estimates, we will assign its value. Using \eqref{deltapulse} in \eqref{pauliequ} yields
\begin{equation}\label{eq:paulinew}
i \hbar \frac{\partial}{\partial t} \psi(t,\vec{x}) = 
 \left(
-\frac{\hbar^2}{2m} \, \triangle +\mu\, \tau\,
\delta(t) \,\vec{\sigma} \cdot
\vec{B}(\vec{x}) \right) \psi(t,\vec{x}).
\end{equation}
Since for $t \neq 0$, equation \eqref{eq:paulinew} is identical to the free Schrödinger equation, we write the wave function as the sum of a free evolution before and after the pulse and try to find the transition conditions through the pulse.  
 Accordingly we write the wave function as
\begin{equation}\label{eq:paulipsi}
\psi(t,\vec{x})=\Theta_+(t)\,\psi_+(t,\vec{x})+\Theta_-(t)\,\psi_-(t,\vec{x})
\end{equation}
with $\Theta_{+}$  representing the Heaviside step function and $\Theta_{-}:= 1 - \Theta_+$.
 Using (\ref{eq:paulipsi}) in (\ref{eq:paulinew}), we obtain
\begin{multline} 
\Theta_+ (t)\left( i  \frac{\partial}{\partial t} +\frac{1}{2m} \, \triangle  \right) \psi_+  + \Theta_-(t) \left( i \frac{\partial}{\partial t} +\frac{1}{2m} \, \triangle \right)\psi_- \\ \label{eq:paulinew1}
+\,i\, \delta(t) \Big(  \psi_+  -\psi_-\Big)  - \,\vec{\sigma} \cdot
\vec{B}\,\, \delta(t)\, \Big(\Theta_+(t) \psi_+ +\Theta_-(t) \psi_- \Big) =0,
\end{multline}
where we have omitted the coordinate dependence of $\psi$ and $\vec{B}$ and set $\hbar=\tau=\mu=1$. Requiring that both $\psi_+$ and $\psi_-$ satisfy the free  equation, the terms in the first line of \eqref{eq:paulipsi} vanish.

The interaction term contains mathematically ill-defined expressions of the form $\Theta_+\delta$. However, the theory of Colombeau generalized functions provides both a rigorous mathematical formalism to handle products of arbitrary distributions   and  also the application-oriented concept of association  for calculations on the distributional level without making use of the full theory.\footnote{Two generalized functions $[(f_{\varepsilon})_{\varepsilon \in (0,1]}]$ and $[(g_{\varepsilon})_{\varepsilon \in (0,1]}]$, these are  (equivalence classes of) families of smooth functions, are associated to each other, if they are equal in the distributional limit,  meaning that
$\lim_{\varepsilon\rightarrow 0}\int_{\mathbb R} \big(f_{\varepsilon}(t)-g_{\varepsilon}(t)\big) \varphi(t) \mathrm dt=0$ for all functions $\varphi$ from a certain test function space.} All we need to know here is that 
the product $\Theta_+\delta$ (which can be defined in a precise way as a Colombeau generalized function) is associated to 
$ a \delta$, where $a$ is a yet undetermined constant, i.e.\ $\Theta_+\delta = a \delta$ in the sense of association. 
 Employing this relation in equation \eqref{eq:paulinew1}, we obtain 
\begin{equation}\label{eq:matching_cond}
\delta(t)\left( \big( \mathbb I +i\,a\,\,\vec{\sigma} \cdot \vec{B} \,\big) \psi_+ -\big(  \mathbb I -i\,(1-a)\,\vec{\sigma} \cdot \vec{B}\,\big) \psi_- \right) = 0,
\end{equation}
where $ \mathbb I$ is the unity matrix and the term in parentheses has to be evaluated at $t=0$. Following \citet{Ba97}
and \citet{BaAi:18},
 we require  that the Pauli (Schrödinger) probability density and current
satisfy the continuity equation  
$\dot{\rho\,}+\vec{\nabla}\cdot \vec{j}=0$.
It is easy to see that $\rho=\psi^{\dagger}\psi$ is conserved across the pulse, yielding the additional condition\footnote{In a non-distributional setting, a solution of the Pauli  equation automatically satisfies the continuity equation. In our case however, it has to be imposed separately, since, in general, equality in the sense of association  is broken in non-linear operations.}
\begin{equation*}\label{eq:pauliconti}
\delta(t)\,\left(\psi^{\dagger}_+ \psi_+ - \psi^{\dagger}_- \psi_- \right)= 0
\end{equation*}
which together with  \eqref{eq:matching_cond} leads to
\begin{equation*}\label{eq:pauliconti}
 \frac{1+\big(1-2a(1\!-\!a)\big)|\vec B(0,\vec x)|^2+a^2(1\!-\!a)^2|\vec B(0,\vec x)|^4}{(1+a^2 |\vec B(0,\vec x)|^2)^2}=1,
\end{equation*}
implying that $a=\frac{1}{2}$.
With this, condition \eqref{eq:matching_cond}  becomes
\begin{equation}\label{eq:paulinew2}
\big( \mathbb I +\tfrac{i}{2}\,\vec{\sigma} \cdot \vec{B}(\vec{x}) \,\big) \Phi_+(\vec{x})=\big(  \mathbb I -\tfrac{i}{2}\vec{\sigma} \cdot \vec{B}(\vec{x})\,\big) \Phi_-(\vec{x})
\end{equation}
where $\Phi_-(\vec{x}):=  \psi_- (0,\vec{x})$ and $\Phi_+(\vec{x}):=  \psi_+(0,\vec{x})$.
Noticing that $(\vec{\sigma}\cdot \vec{B})^2=|\vec B|^2$, it is easy to verify that
\begin{equation*}
\big( \mathbb I +\tfrac{i}{2}\,\vec{\sigma} \cdot \vec{B} \,\big)^{-1}=\frac{\mathbb I -i\frac{1}{2}\vec{\sigma} \cdot \vec{B}}{1+\frac{1}{4} |\vec B|^2}
\end{equation*}
and applying this inverse matrix to equation (\ref{eq:paulinew2}), we end up with the matching condition
\begin{equation}\label{eq:paulinew3}
\Phi_+=\frac{(1- \frac{1}{4}    
|\vec B|^2)\,\mathbb I -i\,\vec{\sigma}\cdot \vec{B} }{1+\frac{1}{4} |\vec B|^2}\,\Phi_-.
\end{equation}

Thus the pulse transforms the incoming data 
$\Phi_-(\vec{x})\equiv  \psi_- (0,\vec{x})$ into outgoing data $\Phi_+(\vec{x})\equiv  \psi_+(0,\vec{x})$.
 In order to obtain the solution after the interaction, one has to solve the free Pauli equation with initial data
 $\Phi_+(\vec{x})\equiv  \psi_+(0,\vec{x})$.

For the free Pauli (i.e.\
Schrödinger-) equation, a plane wave solution with particle momentum $\vec{p}$ is given by
\begin{equation}\label{eq:pwpauli}
\psi_{\vec{p}\,}(t,\vec{x})=\chi \,\,e^{-i\big(\frac{\vec{p}^2}{2m}t -\vec{p} \cdot \vec{x}\big)}
\end{equation}
 where the Pauli spinor $\chi$ corresponds to the polarization of the neutrons and can be chosen arbitrarily.

If we start with a plane wave solution \eqref{eq:pwpauli}  and consider a homogeneous magnetic field $\vec{B}=$const.\ (i.e.\ $ \vec{\mathcal B}(t)=\tau\delta(t)\vec B)$, then the transition condition \eqref{eq:paulinew3} reduces to
\begin{equation}\label{eq:paulimatchxi}
\psi_+(0,\vec{x})=\frac{\big(1-\frac{1}{4}|\vec B|^2 \big)\,\mathbb I -i\,\vec{\sigma}\cdot \vec{B}}{1+\frac{1}{4} |\vec B|^2 }\,\,\chi_-\,\,e^{i\vec{p} \cdot \vec{x}},
\end{equation}
where $\chi_-$ ist the initial polarization of the neutrons. The initial condition \eqref{eq:paulimatchxi} implies that
\begin{equation*}
\psi_+(t,\vec{x})=  \chi_+ \,\,e^{-i\big(\frac{\vec{p}^2}{2m}t -\vec{p} \cdot \vec{x}\big)}
\end{equation*}
where 
\begin{equation}\label{eq:paulimatchxi2}
\chi_+=\frac{\big(1-\frac{1}{4}|\vec B|^2 \big)\,\mathbb I -i\,\vec{\sigma}\cdot \vec{B}}{1+\frac{1}{4} |\vec B|^2 }\,\,\chi_-
\end{equation}
is the final polarization of the neutron.

The particle momentum $\vec{p}$ remains unchanged
since a homogeneous magnetic pulse 
 will only affect the spin of the neutrons. Evoking the  classical analog,
 a homogeneous magnetic field does not exert a resulting force on a magnetic dipole whose magnetic moment  is not aligned with the magnetic field, but a torque.
 \,\\

 As a simple example we consider the initial neutron beam to be polarized in  $z$-direction, i.e.\ 
 $\chi_-=\chi^{\uparrow}$, where $\chi^{\uparrow\downarrow}$ are eigenvectors of the Pauli spin matrix $\sigma_z$, 
and the magnetic field to be orthogonal to the spin polarization, i.e.\ to lie in the $xy$-plane. 
  Then, setting $\vec{B}=(B_x,B_y,0)^T=$const.\ and calculating $\chi_+$ according to (\ref{eq:paulimatchxi2}) yields
\begin{equation}\label{finalstate}
\chi_+ =
 \,\frac{1-   \frac{1}{4}   B^2 }{1+\frac{1}{4}  B^2}\,
\chi^{\uparrow} \\+ \, \frac{ B_y - i\, B_x}{1+\frac{1}{4} B^2}
\chi^{\downarrow}
\end{equation}
where 
${B^2 } = B_x^2 + B_y^2$. From  expression \eqref{finalstate}
 one obtains the ``spin-flip'' probability 
\begin{equation*}
|\langle\chi^{\downarrow}| \chi_+ \rangle|^2 =
 \, \frac{ B^2 }{(1+\frac{1}{4}  B^2)^2}.
\end{equation*}

 \section{Expectation values and numerical estimates}
 In order to obtain numerical estimates, we have  to uncover the constants $\hbar, \mu$ and $\tau$. This amounts to replacing $B_{(j)}$ everywhere by $\frac{\mu}{\hbar} \tau B_{(j)}$, where $\tau$ was  introduced  in \eqref{deltapulse} and $\mu$ is the absolute value of the magnetic moment of the neutron.  As mentioned in the introduction, one way to produce a short intense magnetic pulse is to use an ultrashort pulse laser, since the electric field does not interact with a neutron (at least in the non-relativistic regime). Although we have assumed that the neutrons interact  only at one instant with the pulse, even the shortest laser pulse is of finite duration. We thus identify $\tau$ with the duration of the pulse and take  $B_j$ to be the effective magnetic field strength in direction $j$, which we define as the average of the magnetic field strength over the period $\tau$, i.e.\
 $$
 B_j=\frac{1}{\tau}\int\limits_0^{\tau} \mathcal B_j(t)\mathrm dt.  
 $$ 
We will calculate expectation values for the spin operators $S_j:=\frac{\hbar}{2}\sigma_j$.  interaction.
 If the incoming neutron beam is initially polarized along the $z$-axis, i.e.\ 
    $ \langle \chi_-| S_z | \chi_-\rangle =\frac{\hbar}{2}$,
then the expectation value after the interaction with the magnetic pulse is obtained from \eqref{finalstate},
\begin{equation}
\langle S_ z\rangle_{\chi_+} =
 \frac{\hbar}{2} \,\frac{(1-\frac{1}{4}(\frac{\mu}{\hbar}\tau B)^2 )^2-(\frac{\mu}{\hbar} \tau B)^2}{(1+\frac{1}{4}(\frac{\mu}{\hbar}\tau B)^2 )^2}.
 \end{equation}
Assuming that the magnetic field of the pulse is oriented in $x$-direction, i.e.\ 
$$B_y=B_z=0,\quad B_x=B,$$
 condition \eqref{finalstate} yields
\begin{eqnarray*}
\langle S_x\rangle_{\chi_+}&=&0,\\
\qquad\langle S_y\rangle_{\chi_+}&=&
-\frac{\hbar}{2}\,\frac{2\frac{\mu}{\hbar} \tau B (1-\frac{1}{4}(\frac{\mu}{\hbar}\tau B)^2)}{(1+\frac{1}{4}(\frac{\mu}{\hbar}\tau B)^2)^2},
 \end{eqnarray*} 
 which is consistent with the classical analog, since in this situation the torque $\vec M =\vec \mu \times \vec B$ would point in negative $y$-direction and thus the Bloch vector $(\langle S_x\rangle,\langle S_y\rangle,\langle S_z\rangle)^T$, initially pointing in $z$-direction, should rotate in negative $y$-direction. Plots of the expectation values $\langle S_y\rangle_{\chi_+}$ and $\langle S_z\rangle_{\chi_+}$ are shown in Figure 1.
 \begin{figure}
 \center
 \includegraphics[scale=1]{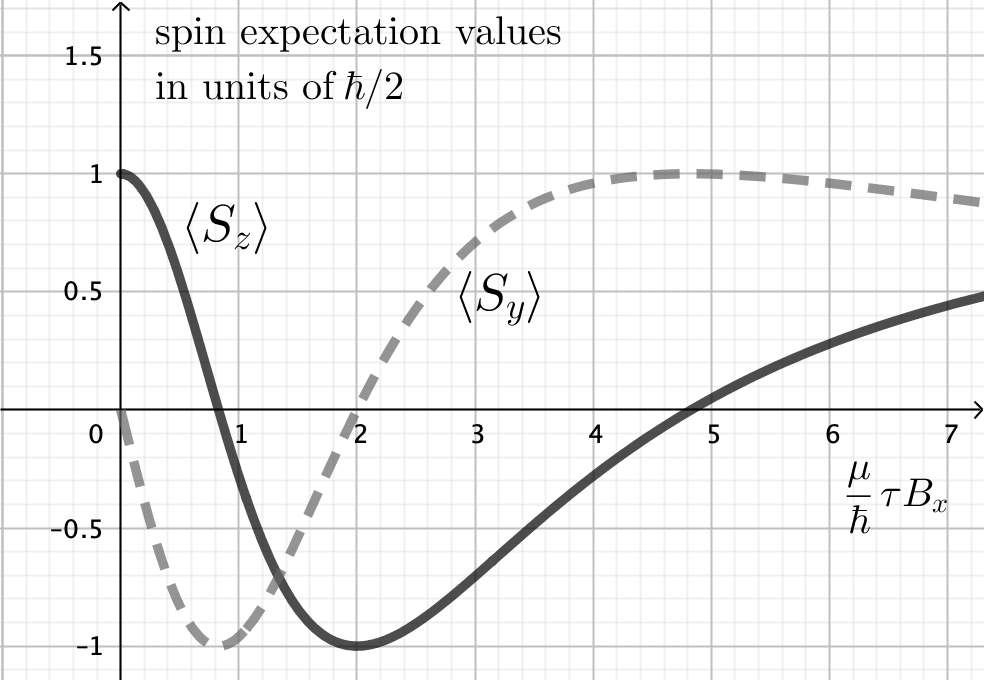}
 \caption{Spin expectation values after the interaction.}
 \end{figure}
For weak fields we can approximate $\langle S_ y\rangle_{\chi_+}$ by
\begin{equation*}\label{eq:larmor}
\langle S_ y\rangle_{\chi_+} \approx  \frac{\hbar}{2}\,\left(-\frac{\mu B_x}{\hbar / 2} \right)\tau= -\frac{\hbar}{2}\,\omega_L\, \tau,
\end{equation*}
 where $\omega_L>0$  is  the Larmor frequency, corresponding to the classical analog of  a magnetic moment precessing with an angular frequency $\omega_L$ in the magnetic field of the pulse. 
 
The shortest laser pulses that can be produced today are not those with the highest intensity. However, laser pulses with a duration $\tau \approx 20\,$fs and a peak intensity   $I\approx 5.5\cdot \times 10^{22}\,\mathrm{W/cm^2}$ have been achieved with the J-KAREN-P laser at KPSI  (Japan), see \citet{Yo:19}. By means of the Poynting vector, this corresponds to  a peak magnetic field strength in the order of   
$$ B\approx \sqrt{\frac{\mu_0 I}{c}}\approx 1.5\cdot 10^{6}\,\mathrm{T}.$$
However, the average value of the magnetic field strength over the volume of the pulse is much smaller. Assuming that the pulse has a total energy $E=5\,$J, a cross-sectional area  $A=2\,(\mu\text{m})^2$ and a duration $\tau=20\,$fs, a crude estimate of the relevant characteristic quantity in terms of the quadratic mean of the magnetic field strength gives
$$ \frac{\mu\tau}{\hbar}B\approx\frac{\mu\tau}{\hbar}\sqrt{\frac{\mu_0 E}{2c\tau A}}\approx  0.3
$$ 
leading to rough estimates for the achievable order of magnitude of 
 the spin-flip probability 
 \begin{equation*}
|\langle\chi^{\downarrow}| \chi_+ \rangle|^2 \approx 0.09 
 \end{equation*}
 and the expectation values 
 \begin{equation*}
\langle S_ z\rangle_{\chi_+} \approx  0.83\,\, \frac{\hbar}{2},\,\ \langle  S_ y\rangle_{\chi_+}\approx  -0.56 \,\, \frac{\hbar}{2}.
 \end{equation*}
 
Given that the assumed conditions can be  accomplished experimentally, our analysis shows that the effect is in principle measurable.

\section{Relativistic corrections and outlook}

   

The Pauli equation is restricted to describe non-relativistic particles with spin. So one may ask whether our results would differ substantially by taking into account relativistic effects.
By the following argument it is possible to estimate the first order corrections in $v/c$.
The above derivation assumes the neutrons to be moving with non-relativistic velocities as viewed by an observer in the  lab system. 
The analysis shows that the momentum of the neutrons not only remains constant through the pulse, but also does not influence the spin interaction. All that matters is the relative orientation of the spin to the magnetic field of the pulse.
This is not so in the relativistic case where in addition, the direction of the velocity with respect to the orientation of the spin and the propagation direction of the pulse becomes  relevant. For example, assuming that the neutron's velocity is orthogonal to the spin orientation, then the electric field  of the pulse will lead to an extra magnetic field strength acting in the neutron's rest frame. This term is of order 
 $(v/c)\frac{\mu \tau}{\hbar}  E 
= \frac{\mu \tau}{\hbar} vB$ and will enter linearly in the expectation values.  

A full relativistic treatment would require a formulation in terms of the Dirac equation. In a further work we intend to give a detailed relativistic analysis for neutrons interacting with an electromagnetic pulse. This would then also allow to discuss the interaction of relativistic neutrons with extremely short and intensive lasers. 

\section*{Declaration of competing interest}
The authors declare that they have no known competing financial
interests or personal relationships that could have appeared to influence
the work reported in this paper.

\bibliography{ABS}

\bibliographystyle{model2-names.bst}

\end{document}